\documentstyle[12pt,epsfig]{article}
\author{Juli\'an Candia$^{a}$, Esteban Roulet$^b$ and  Luis N. Epele$^a$\\
$^a${\it Departamento de F\'{\i}sica, Universidad Nacional de La Plata, 
CC67,}\\{\it La Plata 1900, Argentina}\\
$^b$ {\it CONICET, Centro At\'omico Bariloche, Av. Bustillo 9500,}\\
{\it Bariloche 8400, Argentina}}

\title{Turbulent diffusion and drift in galactic magnetic fields and 
the explanation of the knee in the cosmic ray spectrum}

\begin{document}
\maketitle
\begin{abstract}
We reconsider the scenario in which the knee in the cosmic ray spectrum is
explained as due to a change in the escape mechanism of cosmic rays from the
Galaxy from one dominated by transverse diffusion to one dominated by
drifts. We solve the diffusion equations 
adopting realistic galactic field models and
 using diffusion coefficients
appropriate for strong turbulence (with a Kolmogorov spectrum of
fluctuations) and consistent with the assumed
magnetic fields. We show that properly taking into account these
effects leads to a natural explanation of the knee in the spectrum,
and a transition towards a heavier composition above the knee is
predicted.
\end{abstract}
\section{Introduction}
A well established feature of the
full cosmic ray (CR) energy spectrum is that it has a power-law behavior with
a steepening taking place at the so-called knee,
 corresponding to an energy $E_{knee}\simeq3\times 10^{15}$~eV.
Although the knee of the CR spectrum has been known
since more than four decades, none of the numerous models
proposed so far to explain this feature has managed to become broadly accepted.
Some proposals focus on a possible crossover between different
acceleration  mechanisms
below and above the knee \cite{laga,bier,drury}, 
or exploit the possibility of a change in the
particle acceleration efficiency \cite{fich,joki,koba}.  
Other hypotheses include the nuclear photodisintegration at the
sources \cite{kara,nos}, 
the recent explosion of a single source \cite{erly}, and 
leakage from the Galaxy due
to a change in the confinement efficiency of CRs by galactic 
magnetic fields \cite{syro,wdow,ptus}.
Among the latter, Ptuskin et al. \cite{ptus} in particular consider the knee 
as due to a crossover 
from a diffusive regime dominated by transverse diffusion at low 
energies to another dominated
by drifts, i.e. by the Hall (antisymmetric) diffusion, above the knee. 
Since the diffusion coefficients determine the residence time of the
CRs inside the Galaxy, this means that the observed slope of the CR
energy spectrum will differ from the slope of the original average
source spectrum just due to the energy dependence of the diffusion
coefficients. 
Hence, the scenario just mentioned naturally accounts
for a change in the spectral slope, arising from the diverse energy
dependence of the two different diffusion coefficients. 

In Ref.~\cite{ptus} some important simplifying assumptions had to be done to
make the differential equations determining the CR densities more
tractable. In particular, a simplistic spatial dependence of the
regular field and of the diffusion coefficients was adopted, but it
was however found that 
the spectral slope beyond the knee was quite sensitive
to  the magnetic field configuration adopted and
to the spatial distribution of the sources.
 Also, the
expressions for the diffusion coefficients used were actually only
valid in the limit of small turbulence. Anyhow, a feature similar to
the knee was obtained, but to reproduce the correct value of
$E_{knee}$ a magnetic field spectrum flatter than Kolmogorov was
needed. This last requirement resulted from the fact that
the perpendicular diffusion
coefficient (the only one relevant for the determination of 
the CR densities below the knee) was normalized at low energies
(few~GeV) to
the average value resulting from a comparison with  
the results of a simplified leaky box model of the
Galaxy, and was then extrapolated up to $E_{knee}$ using the fact that
its energy dependence is
directly related to the spectral index of the random magnetic fields.

In the present work we want to further elaborate on the proposal of
Ref.~\cite{ptus}, but using more realistic configurations for the
galactic magnetic fields (including e.g. field
reversals, which enhance field
gradients and can hence affect the CR drifts). We also  consider diffusion
coefficients appropriate for strongly turbulent 
magnetic fields, since
strong turbulence from disturbances in the interstellar plasma is
thought  to be the actual picture
of magnetic fields in the Galaxy, and we will directly adopt diffusion
coefficients normalized from the results of numerical simulations
valid in the relevant energy range of the knee and consistent with the
magnetic fields adopted. 

The beauty of this scenario is that in it the presence of a 
break in the spectrum does not require any special assumption, but just follows
from properly taking into account well established properties of the
propagation of charged particles in regular and turbulent magnetic
fields. Moreover, this break naturally happens at the energies at which 
the knee is actually observed.

\section{The diffusion of cosmic rays in the Galaxy}

\subsection{Regular and random galactic magnetic fields}

The Milky Way, as well as other spiral galaxies, is known to have a large-scale 
regular magnetic field 
$\bf{B_0}(\bf{x})$,
which is determined by measurements of the rotation  due to the
Faraday effect of the
polarization plane of radiation from
pulsars and extra-galactic radio sources \cite{ru88,ha01}.
The magnetic lines are observed to follow the spiral arms, but there
is a variety of models to describe 
the regular magnetic field structures. 
Models of magnetic fields of spiral galaxies are distinguished between 
bisymmetric (BSS)  and axisymmetric (ASS),
defined so that BSS (ASS) fields are antisymmetric (symmetric) with 
respect to a $\pi$ rotation.
For a spiral galaxy with two arms, for instance, in BSS models the field 
reverses its sign between the arms,
while in ASS models it keeps the same orientation. 
In the case of the Milky Way, the regular magnetic field in the
galactic disk is predominantly toroidal, being aligned with the spiral 
structure and having reversals
in direction between nearby arms. Hence, the regular field in the disk
is usually described with a BSS model. 
Furthermore, magnetic field models can 
be classified as symmetric (S) or antisymmetric (A) with 
respect to the galactic plane, and the regular disk component is
believed to be   symmetric with respect to the galactic plane. 
The presence of substantial non-thermal radio emission at high galactic latitudes 
points to the existence of an extended corona with non-negligible
magnetic fields. 
This magnetic halo extends possibly up to $10$~kpc from the galactic plane.
Although the orientation of the regular field in the galactic halo is not known, 
the toroidal component appears to be dominant. The magnetic field structure of the halo
may be similar to that of the galactic disk (for instance if its 
 origin is related to a
galactic wind which somehow extends the field disk properties into the
halo),  but it may also be different since it could be
independently generated through a dynamo mechanism.

Taking into account the main features of the regular magnetic field 
in the Galaxy mentioned above, 
we will assume that the propagation region of the galactic CRs 
is a cylinder of radius $R=20$~kpc and height $2H$, with a halo size $H=10$~kpc. 
Moreover, since the pitch angle of the
spiral arms is quite small ($<10^\circ$), it is a good approximation
to consider that the regular magnetic field is directed in the
azimuthal direction.
We will hence adopt cylindrical coordinates $(r,\phi,z)$ throughout,
taking ${\bf{B_0}}=B_0 \hat{\phi}$ and 
 assuming for simplicity  that the system has 
azimuthal symmetry. In passing, notice that a $\phi -$independent
azimuthal field is necessarily divergenceless.  
In order to reproduce the field reversals of the BSS-S model, 
we will adopt for the regular disk component the expression

\begin{equation}
B_0^{disk}(r,z)=B_d {{1}\over{\sqrt{1+(r/r_c)^2}}} {{1}\over{\sqrt{1+(z/z_d)^2}}}
\sin\left({{\pi r}\over{4~{\rm kpc}}}\right){\rm
th}^2\left(\frac{r}{1\ {\rm kpc}}\right) \ \ ,
\label{b0d}
\end{equation}
where $r_c=4$~kpc is a core radius which smoothes out  near the
galactic center the overall $1/r$
behavior, $z_d=1$~kpc is the adopted magnetic disk scale height, and 
the value of $B_d$ is chosen such that $B_0^{disk}(r_{obs},z_{obs})=-2$~$\mu$G 
at our observation point ($r_{obs}=8.5$~kpc, $z_{obs}=0$)\footnote{The
th$^2(r)$ factor multiplying the regular magnetic field components
just insures that there are no singularities in the $r=0$ axis,
something which would otherwise artificially introduce a singular drift
current along that axis (see below).}. 
Notice the minus sign arising from the fact that the local galactic magnetic
field is nearly directed towards the longitude $l \simeq 90^\circ$, 
i.e. in the $-\hat{\phi}$ direction. This regular field component has
reversals for $r=4,$ 8, 12 and 16~kpc.
For the regular halo component, we will consider it to be described either by
a field structure with radial reversals, namely
\begin{equation}
B_0^{halo}(r,z)=B_h {{1}\over{\sqrt{1+(r/r_c)^2}}} {{1}\over{\sqrt{1+(z/z_h)^2}}} 
\sin\left({{\pi r}\over {4~{\rm  kpc}}}\right){\rm
th}^2\left(\frac{r}{1\ {\rm kpc}}\right)  \ \ ,
\label{b0hr}
\end{equation}   
or by a model without radial reversals, as given by
\begin{equation}
B_0^{halo}(r,z)=B_h {{1}\over{\sqrt{1+(r/r_c)^2}}} {{1}\over{\sqrt{1+(z/z_h)^2}}} 
{\rm
th}^2\left(\frac{r}{1\ {\rm kpc}}\right)\ {\mathcal{R}} \ \ ,
\label{b0h}
\end{equation}
where $z_h=5$~kpc is the adopted magnetic halo scale height, 
and $B_h$ is such that 
$B_0^{halo}(r_{obs},z_{obs})=\pm 1$~$\mu$G. The function $\mathcal{R}$
is just unity  for the symmetric halo case, and it is taken as
${\mathcal{R}}(z)={\rm sign}(z)$ in the case of a halo model 
antisymmetric with respect to the
galactic plane.  
 
Eqs.~(\ref{b0d})--(\ref{b0h}) provide simple but yet realistic
galactic magnetic field models, containing the characteristic features
(such as the reversals) which we believe are important for the
determination of the drift effects.

Superimposed to the regular magnetic field structures of the 
galactic disk and the extended halo, 
an irregular magnetic field $\bf{B_r}(\bf{x})$ due to turbulence 
in the interstellar plasma is known to exist, with
the largest eddies having a  scale $L_{max}\simeq 100$~pc. 
This random field component has a strength comparable to
 the field of the regular component, 
namely $\langle|{\bf{B_r}}|\rangle/\langle|{\bf{B_0}}|\rangle\sim1$--3
locally. 
The existing observational data \cite{arms} 
show that the spectrum of inhomogeneities 
may be the same for the density
of the gas and for the magnetic field, and that it is close to 
a Kolmogorov spectrum. 
Hence, the random magnetic field
can be assumed to be given by a power-law spectrum with Fourier
components giving rise to a magnetic energy density d$E/{\rm
d}k\propto k^{-5/3}$, 
for $k\geq2\pi/L_{max}$. 
The random field intensity will be considered to 
possess a smooth spatial dependence analogous to the envelope of
the regular fields assumed above, namely
\begin{equation}
|{\bf{B_{r}}}(r,z)|=B^*_r {{1}\over{\sqrt{1+(r/r_c)^2}}} {{1}\over{\sqrt{1+(z/z_r)^2}}} \ \ ,
\label{breq}
\end{equation}
where $z_r=3$--5~kpc and $B^*_r$ is such that 
$|{\bf{B_r}}(r_{obs},z_{obs})|=2$--4~$\mu$G.
Since we are concerned with highly relativistic particles of speed $v\simeq c$, 
which is much larger than
the Alfv\'en velocity $v_A$ characteristic of the magnetic wave propagation, 
it is well justified to assume the magnetic disturbances to be static, 
and thus to neglect electric fields.  

\subsection{The diffusion equation}

On scales larger than the scale of magnetic irregularities, 
 the cosmic ray transport across the turbulent galactic magnetic fields 
can be well described by means of a diffusion equation,
in which the components of the diffusion tensor depend on both the regular field and
the random turbulence.  
Neglecting nuclear fragmentation and any energy losses, the steady-state diffusion equation
for the cosmic ray density $N(\bf{x})$ reads
\begin{equation}
\nabla \cdot {\bf J}=Q(\textbf{x}) \ \ ,
\label{diffeq1} 
\end{equation}
where $Q$ is a source
term and the CR current is given by 
\begin{equation}
J_i=-D_{ij}(\textbf{x})\nabla_jN(\textbf{x}) \ \ .
\label{jcr}
\end{equation}
In general, the diffusion tensor $D_{ij}$ can be written as
\begin{equation}
D_{ij}=\left(D_{\parallel}-D_{\perp}\right)b_ib_j+D_{\perp}\delta_{ij}+D_A\epsilon_{ijk}b_k \ \ ,
\end{equation}
where ${\bf{b}}={\bf{B_0}}/{|\bf{B_0}|}$ is a unit vector along the regular magnetic field,
$\delta_{ij}$ is the Kronecker delta symbol, and $\epsilon_{ijk}$ 
is the Levi-Civita fully antisymmetric tensor.   
The symmetric terms contain the diffusion coefficients parallel and perpendicular to the
mean field, $D_{\parallel}$ and $D_{\perp}$, which describe diffusion due to small-scale turbulent
fluctuations. The antisymmetric term contains the Hall diffusion
coefficient $D_A$,  which
is responsible for the macroscopic drift currents.
The diffusion equation can be rewritten as
\begin{equation}
-\nabla_iD_{Sij}\nabla_jN+u_i\nabla_iN=Q \ \ ,
\label{diffeq2}
\end{equation}
where $D_{Sij}= \left(D_{\parallel}-D_{\perp}\right) b_ib_j+D_{\perp}\delta_{ij}$ contains
the symmetric part of the diffusion tensor, 
and $u_i=-\epsilon_{ijk}\nabla_k\left(D_Ab_j\right)$
is called the drift velocity.
Notice also that the CR current in Eq.~(\ref{jcr}) may be written as
\begin{equation}
{\bf J}=-D_\perp\nabla_\perp N-D_\parallel\nabla_\parallel N+D_A{\bf
b}\times \nabla N \ \ ,
\end{equation}
where $\nabla_\parallel\equiv {\bf b}({\bf b}\cdot \nabla)$ and
$\nabla_\perp\equiv \nabla-\nabla_\parallel$. Hence, the
macroscopic drift current is orthogonal to the overall direction of
the regular field and to the CR density gradient.

Under the assumption  mentioned previously  that
the regular magnetic field is directed in the
azimuthal direction (i.e. $b_r=b_z=0$, with $b_{\phi}=\pm1$), and that
the system has  azimuthal symmetry 
(${{\partial}/{\partial\phi}}=0$), Eq.~(\ref{diffeq1}) simplifies to
\begin{equation}
\left[-{{1}\over{r}}{{\partial}\over{\partial r}}\left(rD_{\perp}
{{\partial}\over{\partial r}}\right)-
{{\partial}\over{\partial z}}\left(D_{\perp}{{\partial}\over{\partial z}}\right)+
u_r{{\partial}\over{\partial r}}+u_z{{\partial}\over{\partial
z}}\right] N(r,z)=Q(r,z) \ \ , 
\label{difeqcil}
\end{equation}  
where the drift velocities are 
\begin{equation}
u_r=-{{\partial(D_Ab_{\phi})}\over{\partial z}}
\label{ureq}
\end{equation}
and
\begin{equation}
u_z={{1}\over{r}}{{\partial(rD_Ab_{\phi})}\over{\partial r}} \ \ . 
\label{uzeq}
\end{equation}
These expressions for the drift velocities already make apparent the
possible relevance of including field reversals in the regular fields,
since this certainly affects the spatial gradients of the diffusion coefficients.

At the surface
 $\Sigma$ corresponding to the  boundaries of the halo, the CR density
has to become  negligible, 
since particles can
escape freely into the intergalactic space. The boundary conditions that apply to
Eq.~(\ref{difeqcil}) are thus $N|_{\Sigma}=0$. 

Notice that the diffusion treatment of the CR propagation will be
adequate as long as the Larmor radius, which for a
relativistic particle of energy $E$ and charge $Z$ is  given by  
\begin{equation}
r_L\equiv {pc\over Ze|\bf{B_0}|}\simeq{{E/Z}\over{10^{15}~{\rm eV}}}\left({|{\bf{B_0}|}
\over{\mu{\rm G}}}\right)^{-1} {\rm pc}\ \ ,
\label{rleq} 
\end{equation}
is smaller than the typical scale height of the magnetic field. Hence,
for CRs trapped in the disk of the Galaxy, this requires $E/Z<{\rm
few}\times 10^{17}$~eV, and somewhat larger values could result if the magnetic
fields in the halo were sizeable. Clearly the knee region is well inside
this domain.

\subsection{Calculation of transverse and Hall diffusion coefficients}

The components of the diffusion tensor 
that define the CR diffusion process 
through Eqs.~(\ref{difeqcil})--(\ref{uzeq}) 
depend on both the regular field and the 
random turbulence, but
unfortunately a comprehensive theory capable of 
determining the diffusion coefficients 
~from our knowledge about the magnetic fields is lacking.
 
~From quite general considerations concerning the ensemble average of the 
two-time particle velocity correlations, 
i.e. $R_{ij}(t)\equiv\langle v_i(t_0)v_j(t_0+t)\rangle$,
applied to the Taylor-Green-Kubo formula \cite{bieber}, 
the following theoretical relations involving the diffusion coefficients
are expected to hold: 
\begin{equation}
D_{\parallel} = {{v r_L}\over{3}}{\omega\tau_{\parallel}}\ \ ,
\label{dpar}
\end{equation}
\begin{equation}
D_{\perp} = {{v r_L}\over{3}}{{\omega\tau_{\perp}}
\over{1+\left(\omega \tau_{\perp} \right)^2}}
\label{dperp}
\end{equation}
and
\begin{equation}
D_A={{v r_L}\over{3}}{{\left(\omega\tau_A\right)^2}
\over{1+\left(\omega \tau_A \right)^2}}\ \ ,
\label{daeq}
\end{equation}
where $v\simeq c$ is the particle speed 
and $\omega\equiv v/r_L$ is the angular gyrofrequency\footnote{The
sign of $D_A$ in Eq.~(\ref{daeq}) corresponds to the diffusion of
positively charged particles, and the opposite sign will hold for
negatively charged ones.}. It should be noticed
that three different timescales $\tau_{\parallel},\tau_{\perp}$ and $\tau_A$ 
appear here as the effective timescales for the decorrelation of
particle trajectories. Indeed, the trajectory decorrelation due to a given
component of the diffusion tensor will be in general associated to
some particular physical
processes which may not always be contributing to the other components
of the diffusion tensor.
For instance,  for low turbulence levels the 
field line random walk appears as the dominant contribution
to the transverse diffusion, but it plays no 
major role in the parallel diffusion. 

There is however no general theory providing the decorrelation timescales.
In particular, this turns out to be a severe restriction 
for the study of particle diffusion 
under highly turbulent conditions. It should be noticed 
that in treating CR diffusion in
the Galaxy we are precisely concerned with high turbulence levels, not
only because the mean random magnetic field is of the order 
of the mean regular field, but also because 
when there are  reversals
in the orientation of the regular field, there exist regions in 
which the regular field becomes
negligible and hence inside them the turbulence largely prevails.

To go beyond the general expressions in Eqs.~(\ref{dpar})--(\ref{daeq})
requires then to make some simplifying assumptions about the timescales for
decorrelations. For instance, in Ref. \cite{bieber} 
a common decorrelation timescale $\tau$ was assumed to hold 
for both the perpendicular and the
antisymmetric diffusion ($\tau \equiv \tau_{\perp} = \tau_A$). 
It was then assumed that the major contribution to perpendicular diffusion is
due to field line random walk and,  
based on simple considerations, the Ansatz 
\begin{equation}
\omega \tau = \frac{2}{3} \frac{r_L}{D_{flrw}} 
\label{omtau}
\end{equation}
was adopted to relate the decorrelation timescale to the diffusion coefficient
$D_{flrw}$ associated with the field line random walk.
Provided either an analytical or numerical estimate for $D_{flrw}$,
which depends on both the form of the turbulence spectrum 
and the turbulence level assumed, $D_{\perp}$ and $D_A$
are thus obtained. However, the numerical study of Ref. \cite{giaca} shows
that the Ansatz proposed in \cite{bieber} 
leads to results clearly inconsistent with the
simulations, so it may be concluded that Eq.~(\ref{omtau}) 
is not appropriate to model all
effects contributing to the randomization of the particle trajectories.

Expressions similar to Eqs.~(\ref{dpar})--(\ref{daeq}), 
but in which only a single timescale
$\tau$ appears for the three diffusion coefficients, 
were obtained in several previous analytic studies 
that assumed a single scattering process to be responsible for all the
decorrelations. This happens, for instance,     
by considering a nearly isotropic particle distribution under 
an homogeneous scattering process,
in which $\tau$ represents the characteristic time of isotropization~\cite{isen}.
A similar structure was also found for collisional diffusion coefficients in 
a plasma in thermal equilibrium \cite{bales}.    
Furthermore, regarding $\tau$ as the 
scattering time, these expressions can also be obtained by treating the magnetic 
disturbances as hard-sphere scattering centers~\cite{glees}, and is known as the 
``classical scattering result''~\cite{park,chap,form}. 
 
Very recently, Casse et al. \cite{casse} presented 
extensive Monte Carlo measurements of parallel and transverse diffusion of 
relativistic test particles propagating in 
strong stochastic Kolmogorov fields, extending
previous numerical results \cite{giaca}. 
~From these investigations, it turns clearly
out that the extrapolation of the previously mentioned theoretical 
models to account for particle diffusion
in strongly turbulent fields is generally not applicable. 
For instance, the numerical simulations \cite{giaca, casse} 
show that, while Eq.~({\ref{dpar}) for parallel diffusion 
appears to agree well with the obtained results, Eq.~(\ref{dperp}) for transverse 
diffusion (with $\tau\equiv\tau_{\parallel}=\tau_{\perp}$) 
clearly fails, as expected from the fact that field line random
walk is in general not negligible. 

Given these facts, what we will
do instead is to directly extract   
the transverse diffusion coefficient 
~from the numerical results of Ref.~\cite{casse}. These results will
be applicable 
in the regime of interest for the study of the knee region, since the
relevant energy parameter, $\rho\equiv 2\pi r_L/L_{max}$, takes the
value $\rho\simeq 0.1$ for protons near the knee, which is just inside the range
[0.001,10] explored in Ref.~\cite{casse}. 

In Ref.~\cite{casse} $D_\perp$ was obtained 
 as a function of rigidity and
turbulence level (their Figure~5), were the
turbulence level is defined as $\eta\equiv B_r^2/(B_r^2+B_0^2)$. 
It turns out that an approximate expression for $D_{\perp}$,  valid in the range
$0.1\leq\eta\leq 1$, is just 
\begin{equation}
D_{\perp} \simeq 8.7\times 10^{27} e^{3.24\eta} 
\left({{\bar{r}_L}\over{\rm{pc}}}\right)^{1/3} {{\rm{cm^2}}\over{\rm{s}}} \ \ ,  
\end{equation}
where $\bar{r}_L$ is  an effective Larmor radius defined as in
Eq.~(\ref{rleq}) but replacing $|{\bf{B_0}}|
\rightarrow \sqrt{B_0^2+B_r^2}$.
It should be remarked that, although this expression
is obtained from numerical simulations, the dependence on rigidity 
through a power $1/3$ appears also in theoretical expressions (valid in
the limit of small turbulence) under the assumption of a Kolmogorov
spectrum for the random fields \cite{ptus}.    

Regarding the drift effects,  there are at  present no numerical
studies  concerning the Hall 
diffusion under turbulent conditions, so that some additional
assumptions had to be made. 
Notice first that from Eq.~(\ref{daeq}) we can obtain the value of
$D_A$ in the two limiting cases of very strong or weak turbulence. In the
first case, i.e. for $\omega\tau_A\to 0$,  one has as expected that
$D_A\to 0$, so that drift effects become small (this limit is of 
particular interest in the regions with
field reversals, in which the turbulence will tend to suppress the
drift effects). On the other hand, in the limit of very weak turbulence
($\omega\tau_A\gg 1$), the obtained expression 
tends to the well-known result $D_A= vr_L/3$ 
of the quasi-linear approximation. In order to obtain a smooth
transition between the two limiting regimes, and lacking theoretical
or numerical results for $\tau_A$, we will just adopt Eq.~(\ref{daeq})
for $D_A$ but assuming here a common timescale for all decorrelation
processes. Under the 
assumption of a common decorrelation timescale, it turns out in
particular that
\begin{equation}
{{D_{\parallel}}\over{D_{\perp}}} = 1 + \left(\omega \tau \right)^2 \ \ ,
\label{dpdp}
\end{equation} 
and this ratio was obtained as a function of rigidity and turbulence
level in Ref.~\cite{casse} (see their Figure~6). The ratio
$D_{\parallel}/
D_{\perp}$ turns out to
be almost independent of the energy in the range studied, 
and using Eq.~(\ref{dpdp}) we find that the simple approximate relation
$\omega\tau\simeq4.5 (|{\bf{B_0}}|/|{\bf{B_r}}|)^{1.5}$  reasonably reproduces the
numerical results for $D_{\parallel}/D_{\perp}$. This result for
$\omega\tau$ may now be replaced 
 in Eq.~(\ref{daeq}) (after changing $\tau_A\rightarrow \tau$)
and this provides  an expression for the Hall diffusion coefficient. 
The resulting expression for $D_A$ has the proper
behavior in the limits of very strong or very
weak turbulence, and the criterium adopted to obtain $\omega\tau$ 
just allows to get a smooth transition between the
two regimes. We do not expect the general results obtained to
change qualitatively if a different prescription to regulate $D_A$ were
adopted. 
Anyhow, further investigations concerning the Hall diffusion under turbulent
conditions are currently under progress. 

\section{Results and comparison with observations}
 
We have assumed the regular galactic magnetic field to be composed of a disk component
with radial reversals and symmetric with respect to the galactic plane
(given by Eq.~(1)),  
and a halo component which could be described either by a field
structure with radial 
reversals (Eq.~(2))  or by a model without radial reversals
(Eq.~(3)). In the former case,  
 it is supposed that the halo is an extension of the disk magnetic
field, and hence it 
will be considered to be also symmetric with respect to the galactic
plane and directed  
parallel to the disk component throughout the Galaxy. 
In contrast, in the latter case all relevant variants concerning symmetry and orientation 
will be explored, since this case may correspond to an independently generated halo field.
Hence, in this case four different configurations arise from choosing the orientation
(through the sign of $B_h$) and the symmetry with respect to the galactic plane (through  
the function $\mathcal{R}$).
     
Adopting a certain galactic field model 
and assuming a given source distribution $Q(r,z)$, the CR diffusion is determined
by solving the diffusion equation given by Eqs.~(\ref{difeqcil})--(\ref{uzeq}). 
Following Ref.~\cite{ptus}, 
the sources are assumed to lie on a thin disk of height $2h_s=400$~pc,
so that  the source term takes the form 
$Q(r,z)=q(r)\theta(h_s-|z|)$, where $\theta$
is the Heaviside function.   
Concerning the radial source distributions, 
both localized\footnote{These sources actually correspond to rings around
the galactic center, since a non-central 
point-like source would break the azimuthal
symmetry assumed.} (i.e., $q(r)\propto \delta(r-r_s)$, with $r_s=3$
and 6~kpc) 
and extended ($q(r)=const.$) distributions have been studied.   
The resulting differential equation has been solved numerically 
using the Crank-Nicolson discretization and
the Liebmann method \cite{press}.

\begin{figure}
\centerline{{\epsfxsize=5.8truein \epsffile{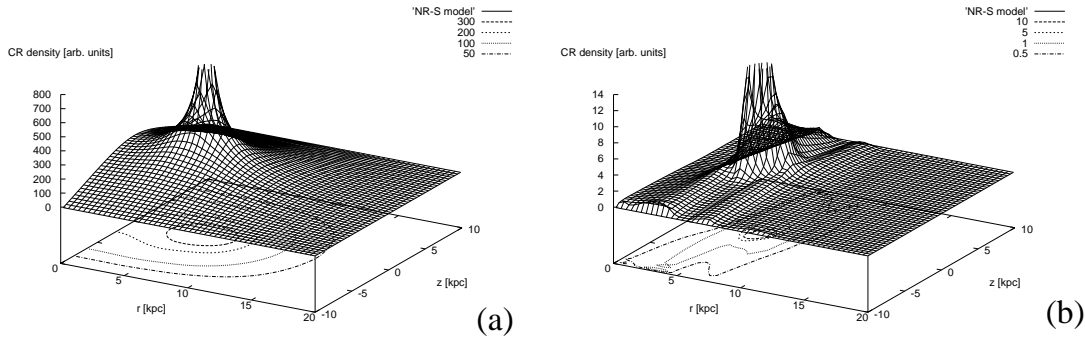}}}
\caption{Spatial distributions of the CR density within the Galaxy, 
assuming a halo without radial reversals 
and symmetric with respect to the galactic plane, 
for a source localized at $r_s=6$~kpc. 
Notice that the sharp peaks occurring around the source location
have been truncated.
(a) At low energies below the knee 
the CR particle diffusion is mainly due to small-scale turbulent fluctuations (governed by $D_\perp$).
(b) At high energies above the knee, the macroscopic drift currents lead to
significant large-scale asymmetries in the CR density distribution.}
\label{fig1}
\end{figure}

In order to illustrate the role of drifts in the CR diffusion, 
let us first consider a localized source at $r_s=6$~kpc, and 
adopting for the random field an amplitude
$|{\bf{B_r}}(r_{obs},z_{obs})|=2$~$\mu$G and a scale height $z_r=3$~kpc.
Figures~\ref{fig1}(a) and (b) show the spatial distribution
of the CR density within the Galaxy, assuming a halo without radial reversals 
and symmetric with respect to the galactic plane,
for energies both below and above the knee. As expected from the
preceding  discussions, 
it is observed that at low energies the CR particle diffusion is mainly due to 
small-scale turbulent fluctuations (governed by $D_\perp$),
while at higher energies the bulk of the CRs is driven according 
to the macroscopic drift currents,
leading to significant large-scale asymmetries in the CR density distribution.    

The influence of the drift effects on the CR spectrum at our observation point depends on 
the galactic field model, on the particular set of parameters considered within the given model 
(i.e. the field amplitudes and scale heights assumed for both the regular and 
random field components), and on the source distribution. Notice that
the relevant parameter in the diffusion coefficient is the Larmor
radius of the CRs, implying that the impact of drift effects will depend
on the CR rigidity, and hence on the value of $E/Z$.

Figure~\ref{fig2} shows the normalized CR flux $\phi$ as a function of $E/Z$ 
for the different regular galactic field models described in Sect. 2.1
with the same source and local random field amplitude 
as in Figure~1. The random field scale height considered
is $z_r=5$~kpc. Notice that in this and the following figures, the CR flux
appears multiplied by $E^{\gamma_Z+1/3}$, where $\gamma_Z$ is the spectral index corresponding
to the component of charge $Z$ at the source (see Eq.~(20) below). Since transverse diffusion
depends on rigidity through a power $1/3$ (see Eq.~(18)), it turns out that $\phi E^{\gamma_Z+1/3}$
is flat at low energies well below the knee.  
$R-S$ denotes the halo with radial reversals and symmetric with respect to the galactic plane,
while $NR-S(A)$ refers to the halo without radial reversals and symmetric 
(antisymmetric) with respect to 
this plane. Furthermore, the orientation of the halo relative to the regular disk component is 
distinguished by means of the sign of $B_d \cdot B_h$.

\begin{figure}
\centerline{{\epsfysize=2.5in \epsffile{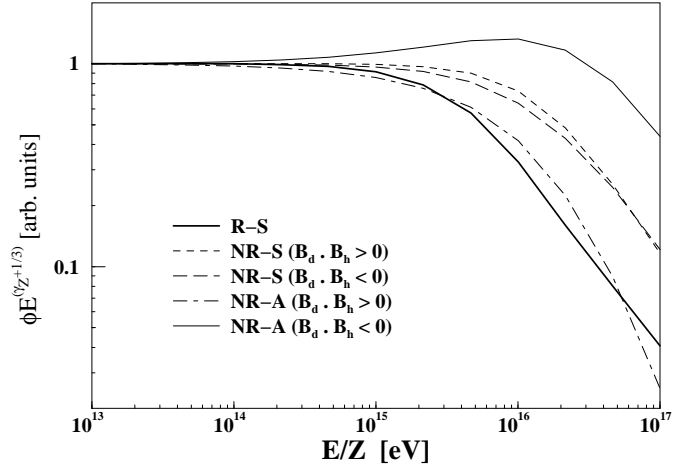}}}
\caption{Plots of the normalized differential energy spectrum versus $E/Z$ 
for the different regular galactic field models investigated.
$R-S$ denotes the halo with radial reversals and symmetric with respect to the galactic plane,
while $NR-S(A)$ refers to the halo without radial reversals and symmetric 
(antisymmetric) with respect to this plane. Furthermore, 
the orientation of the halo relative to the regular disk component is 
distinguished by means of the sign of $B_d \cdot B_h$, as indicated.}
\label{fig2}
\end{figure}

It can be seen that most of the models considered show the appropriate
tendency to account for the knee, since a steepening in the spectrum
appears at the right value of $E/Z$ (i.e. for $E\simeq E_{knee}$ for
protons, and higher for heavier nuclei). Drift effects are
particularly enhanced for the antisymmetric halo case, as could be
anticipated from Eq.~(\ref{ureq}) since for $b_\phi\simeq \pm {\rm
sign}(z)$ one has $u_r\simeq \mp 2D_A\delta(z)$ on the galactic plane, 
and this additional contribution to the drifts significantly affects 
the results. Moreover, the halo contribution to the vertical drift
velocity $u_z$ will be converging (or diverging) from both sides of
the galactic plane (rather than being similarly oriented as in the
symmetric models).
One obtains in this case a more pronounced knee for $B_d\cdot B_h>0$, while 
a flattening of the spectrum just before the knee is 
observed for $B_d\cdot B_h<0$. 
The pronounced bump observed in Figure~\ref{fig2} for this case 
is in disagreement with the observed spectra, so
that this model would be observationally disfavored. However, 
the possible existence of a  smaller bump just below the knee in the
actual data was noticed by Kempa et al. \cite{kempa} many years ago,
although it still remains
somewhat controversial. It is interesting however that features of this
kind may arise from drift effects in some particular configurations of
regular galactic magnetic fields.

\begin{figure}
\centerline{{\epsfysize=2.5in \epsffile{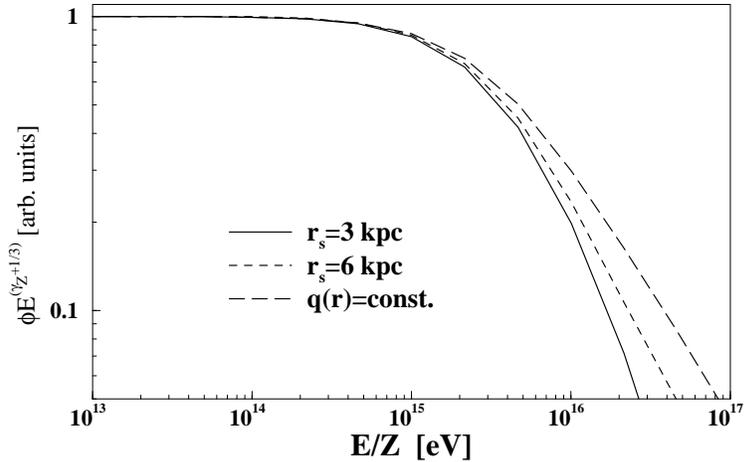}}}
\caption{Plots of the normalized differential energy spectrum versus $E/Z$ 
for the $R-S$ case and different source distributions, namely homogeneously 
extended ($q(r)=const.$) and localized sources ($q(r)\propto \delta(r-r_s)$, with $r_s=3$
and 6~kpc).} 
\label{fig3}
\end{figure}

Figure~\ref{fig3} shows the $R-S$ case for an extended source distribution 
$q(r)=const.$, as 
well as for sources localized at $r_s=3$~kpc and $r_s=6$~kpc, with the
same   random field as in Figure~\ref{fig1}.
As may be expected, the drift effects appear stronger the farther the
source is located, since drift effects can remove high energy CRs from
the galactic plane all the way from the source to the observer.
The same qualitative behavior has
also been observed in the other cases studied.

\begin{figure}
\centerline{{\epsfysize=2.5in \epsffile{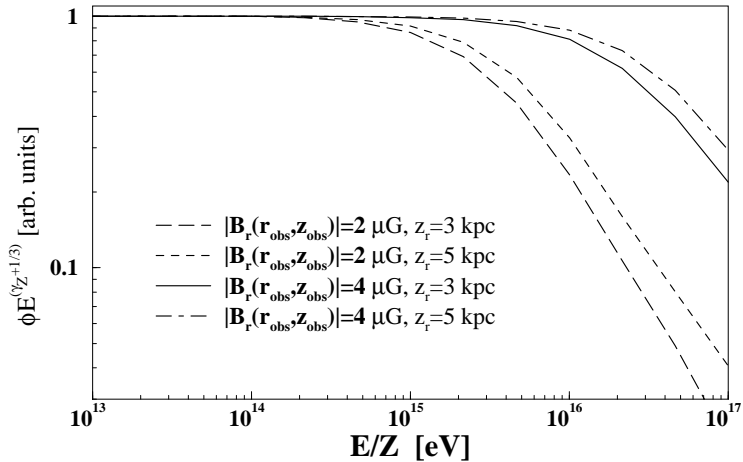}}}
\caption{Plots of the normalized differential energy spectrum versus $E/Z$ 
for the $R-S$ case and a source localized at $r_s=6$~kpc. 
The random field parameters take the values $|{\bf{B_r}}(r_{obs},z_{obs})|=2$ 
or 4~$\mu$G and $z_r=3$ or 5~kpc, as indicated.}
\label{fig4}
\end{figure}

Figure~\ref{fig4} displays the effects of varying the parameters relevant to the random
field. The data correspond to the $R-S$ case and the random field parameters take
the values $|{\bf{B_r}}(r_{obs},z_{obs})|=2$ or 4~$\mu$G and $z_r=3$
or 5~kpc. As expected, 
the drifts are clearly stronger when
the random field is suppressed, either by decreasing its overall amplitude or its scale
length. Again, the same qualitative behavior has
been observed in the other relevant field configurations, and is due
to the increased suppression of $D_A$ under stronger turbulence conditions.

To compare our results with experimental data, it is assumed that
the differential fluxes of different nuclei  emitted by the sources
have 
power-law distributions, i.e.
\begin{equation}
\phi_Z^0(E)=\Phi_Z^0E^{-\gamma_Z}\ \ ,
\end{equation} 
where $1\leq Z\leq 26$.
The intensities $\Phi_Z^0$ and spectral indices $\gamma_Z$
were taken from the detailed knowledge about CR mass composition below the
knee (and for energies per particle above $\sim  few \ \ Z \times
10^{10}$~eV) \cite{wiebel}, and they were then extrapolated to
energies above the knee. Making the convolution of the spectrum of
each component of charge $Z$ with the modulation effects produced by
the drifts discussed previously and 
summing over
all components from hydrogen to iron, we arrive at the all-particle CR flux given by
the model, which can then be compared with the observed spectra measured by different experiments
(KASCADE, CASABLANCA, DICE, PROTON satellite, and Akeno)~\cite{swordy1,fowler,swordy2,kasc1,kasc2}.

\begin{figure}
\centerline{{\epsfysize=2.5in \epsffile{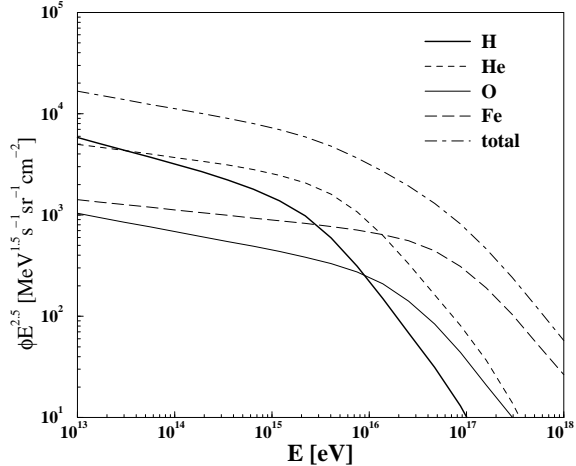}}}
\caption{Plots showing the main contributions to the total CR flux coming 
from protons and nuclei of
helium, oxygen and iron, calculated in the $R-S$ case for a source 
localized at $r_s=6$ kpc.}
\label{fig5}
\end{figure}

Figure~\ref{fig5} shows the main contributions to the total CR flux coming 
from protons and nuclei of
helium, oxygen and iron, calculated in the $R-S$ case for a source 
localized at $r_s=6$ kpc, with 
the random field parameters as in Figure~\ref{fig1}.
Figure~\ref{fig6}(a) shows the total CR flux corresponding to this case
compared with some observations, and clearly a very satisfactory  
 agreement is found between the predictions and the actual
experimental data.
As discussed above and shown in preceding figures,
varying model parameters and considering other galactic field 
models enables to obtain
somewhat different spectra. 
We do not attempt here to make a detailed quantitative test of the
different models, both because the different data still have sizeable
statistic (and systematic) uncertainties, and because the predictions
also rely on several simplifying assumptions. For instance,  the
transport equation was regarded as purely diffusive, neglecting  
spallation of nuclei in the interstellar medium or  convection due to
the possible existence of a galactic wind.
Also, the spatial dependence of the various components of the magnetic
field in the Galaxy were taken as
smooth as possible  in order to simplify the integration of the 
differential equations determining the CR densities. Although the
adopted models are plausible and capture the main expected features of
the galactic magnetic fields, certainly the real ones will be less idealized. 
Moreover, a (sensible) prescription for dealing with the Hall diffusion
under turbulent conditions
had to be  adopted, due to   the lack of appropriate numerical studies 
concerning this issue at present.  
Hence, an accurate fit to the experimental data is not to be expected, 
although it is 
reassuring to observe that the predictions within this scenario tend
to exhibit a remarkable agreement with observations. 

\begin{figure}
\centerline{{\epsfxsize=5.8in \epsffile{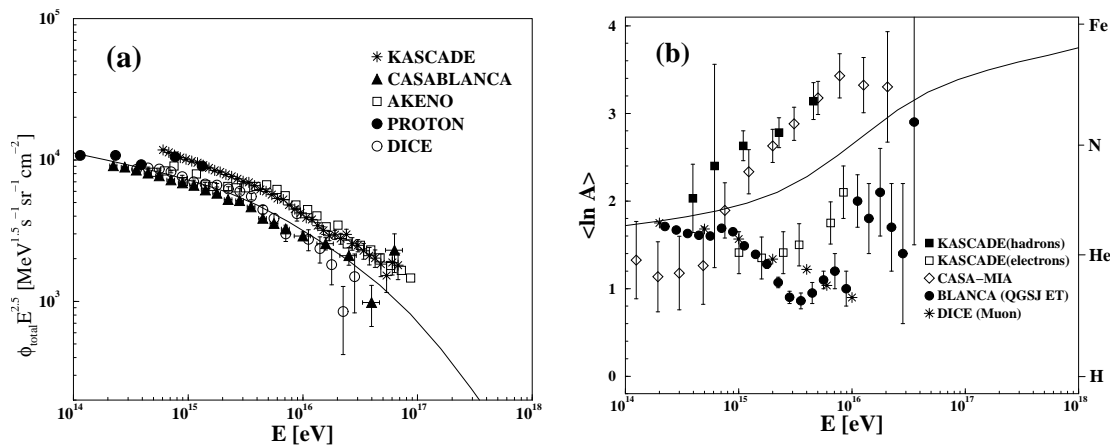}}}
\caption{Comparison between experimental observations and 
model predictions obtained in the $R-S$ case for a source 
localized at $r_s=6$~kpc: 
(a) plot of the total CR flux versus the energy per particle;  
(b) plot of the mean mass composition $\langle \rm{ln} A \rangle$ versus 
the energy per particle.}
\label{fig6}
\end{figure}

Figure~\ref{fig6}(b) shows a plot of the mean mass composition $\langle \rm{ln}
A \rangle$ versus $E$ 
corresponding to the same case as before. For comparison, we also show
the CR mass composition measured by 
different experiments \cite{swordy1,fowler,swordy2,kasc1,kasc2}. 
Since leakage from the Galaxy depends on $E/Z$ 
and is more effective for lighter particles, this scenario can easily
account for a change in the 
composition towards heavier nuclei above the knee, which is apparent
on some of the existing data and in particular in the latest KASCADE
observations \cite{kasc2}.  
However, since above $10^{14}$~eV/nucleus the information
about the CR mass composition has to be obtained indirectly  from 
extensive air shower observations at ground level, still significant
discrepancies are found between the results of experiments using
different techniques.
Indeed, the CR mass composition beyond the knee is not 
definitively established yet, with some observations 
\cite{swordy1} suggesting that it could turn lighter while others
finding that the heavier components become
dominant \cite{kasc1,kasc2}. 
In Figure~\ref{fig6}(b) it can be seen that some data sets suggest that 
the mean mass composition has a dip starting at $E\simeq 10^{15}$~eV, 
while it grows
steadily towards heavier components beyond $E\simeq 3\times
10^{15}$. A dip corresponding to the composition becoming lighter
could possibly result from a bump in the spectrum of the proton
component in the same energy range. The models predict on the
other hand that for increasing energies only the heavier elements
persist, since the lighter ones are efficiently drifted away for
$E/Z\gg E_{knee}$. 

As a summary, we have considered in detail the diffusion and drift
effects of CRs in realistic configurations of the galactic magnetic
fields, adopting diffusion coefficients appropriate for the high levels
of turbulence present in the Galaxy and for a Kolmogorov spectrum of
fluctuations. On the one hand, the larger field gradients associated to
the field reversals present between spiral arms  enhance the drift
effects, but on the other hand a suppression of $D_A$ associated to
the high turbulence present (particularly near field reversals) tends
to reduce the drifts, so that the two effects compete among each
other. We used diffusion coefficients obtained directly through
numerical simulations valid in the relevant energies, and found that
the correct value of $E_{knee}$ can easily be obtained.  Although 
the slope of the spectrum of each nuclear component changes
by a factor $\Delta\alpha\simeq -2/3$ from below to above $ZE_{knee}$,
the envelope of the total spectrum shows a milder change in slope,
which is consistent with the observed spectrum. A progressive
transition towards a heavier composition results in this scenario. 

Certainly a better knowledge of the CR spectrum and especially of the
 CR mass composition around and 
above the knee is of major importance to finally decide which is the correct
 mechanism responsible for this feature.

\section*{Acknowledgments}
Work supported by CONICET, ANPCyT and Fundaci\'on Antorchas, 
Argentina. 

\end{document}